# Achieving Employees' Agile Response in E-Governance: Exploring the Synergy of Technology and Group Collaboration

Ying Bao[1] · Xusen Cheng[2] · Linlin Su[3] · Alex Zarifis[4]

## Abstract

The transformation of technology and collaboration methods driven by the e-government system forces government employees to reconsider their daily workflow and collaboration with colleagues. Despite the extensive existing knowledge of technology usage and collaboration, there are limitations in explaining the synergy between technology usage and group collaboration in achieving agile response from the perspective of government employees, particularly in the e-government setting. To address these challenges, this study provides a holistic understanding of the successful pathway to agile response in e-governance from the perspective of government employees. This study explores a dual path to achieve agile response in e-governance through qualitative analysis, involving 34 in-depth semi-structured interviews with government employees in several government sectors in China. By employing three rounds of coding processes and adopting Interpretative Structural Modeling (ISM), this study identifies the five-layer mechanisms leading to agile response in e-governance, considering both government employee technology usage and group collaboration perspectives. Findings of this study provides suggestions and implications for achieving agile response in e-governance.

✉ Xusen Cheng
  xusen.cheng@ruc.edu.cn

1 School of Economics and Management, China University of Petroleum, Beijing, China

2 School of Information, Renmin University of China, Beijing, China

3 College of Economics and Management, South China Agricultural University, Guangzhou, China

4 Southampton Business School, University of Southampton, Southampton, UK

⁂ Springer

## 1 Introduction

The government is continually expected to deliver high-quality services to citizens and may be criticized for low agility (Nam 2019). The fundamental purpose of governments worldwide has been to achieve agile response in governance; that is, government employees display high effectiveness and efficacy in their work (Lee and Xia 2010). This, in turn, also results in government services for citizens being delivered with effectiveness and efficacy (Janita and Miranda 2018; Santoro et al. 2019). E-government projects, facilitated by Information and Communication Technology (ICT) (Twizeyimana and Andersson 2019), aim to empower government employees, both individually and collaboratively, to deliver higher-quality services to citizens through e-government systems (West 2004). E-government systems, such as government-to-citizen (G2C) and government-to-employees (G2E) systems (Rao 2011), are respectively accessible to government employees and citizens for specific services (Rana and Dwivedi 2015). However, many e-government projects, despite significant investments, fail to meet the expectations to achieve agile governance (Savoldelli et al. 2014; Hooda et al. 2022).

Given the potential benefits of e-government, it is crucial to identify how to successfully achieve agile response through the utilization of e-government systems. A significant number of studies have focused on G2C systems to explain the failure or success of the functionality the systems offer (Gupta and Jana 2003), technology adoption (Venkatesh et al. 2014; Liang et al. 2017), and citizens' satisfaction (Venkatesh et al. 2016). However, there has been limited focus on G2E systems, leading to a gap in understanding how to achieve agile governance from the perspective of government employees. During the COVID-19 pandemic, governments had to rapidly transition to digital workflows, requiring employees to quickly adapt to new platforms for public service management, data processing, and pandemic response coordination. In healthcare sectors, for instance, government employees managed vaccine distribution, tracked cases, and facilitated communication using digital tools. In spite of technology enabled response agility, success government services also relied on effective collaboration between employees or departments. Quick information sharing and responses were essential to handling the logistical challenges of the crisis, highlighting the importance of agility in public administration and accelerating the digital transformation of government.

As a form of digital transformation, e-government projects involve not only the implementation of e-government systems but also the establishment of new workflow that affects the means of collaboration among government employees (Tan and Pan 2003; Verhoef et al. 2021). Therefore, it is imperative to identify the pathway to agile response through both crucial aspects in the setting of e-government: government employee technology usage and government employee group collaboration.

Existing studies have respectively examined factors influencing organizational employees' technology usage intentions (e.g., Gallivan et al. 2005) and their collaboration performance (e.g., Fu et al. 2020). However, they have failed to consider reasons for agile response, which necessitate an integrated view of both G2E systems usage and group collaboration. Agile response in e-governance involves a dynamic interplay between how government employees use G2E systems and how they col-laborate effectively as a group. For instance, extant research has initially explored government employees' technology usage, indicating that factors such as information quality, system quality, and service quality determine their satisfaction, usage intention, and net benefits, drawing on the information systems success model (Stefanovic et al. 2016). Collaboration has also been widely recognized as beneficial in virtual or semi-virtual environments facilitated by ICT (Cheng et al. 2021). Cross-sector interactions and group collaboration among government employees are crucial processes for service delivery in governments (Nam and Pardo 2014), especially in online workspace; however, existing studies have overlooked how group collaboration perceptions can benefit from e-government systems. Additionally, traditional workflow and routines may present significant challenges for government employees in adapting to e-government systems for virtual or semi-virtual collaborations to achieve agility. While prior studies failed to consider the interplay between technology and group dynamics in the context of G2E systems. Thus, there is a call for a holistic understanding of the pathway to agile response in e-governance, considering the two crucial aspects of government employee technology usage and group collaboration. This study aims to address this gap by investigating the dual pathways—technology usage and group collaboration—that contribute to employees' agile response in e-governance, as depicted in Fig. 1.

In terms of the practical requirements and theoretical gaps, this paper provides the following research question:

**RQ** *How does the implementation of e-government systems through a dual path, involving both government employee technology usage and group collaboration, achieve agile response in e-governance?*

To address the research question, this study employed a qualitative methodology, gathering interview data from 34 Chinese government employees who used e-government systems to conduct collaborative tasks. China's government is one of the largest in the world, and the large scale of its government employees presents unique organizational challenges in implementing e-governance (Li and Shang 2023; Shou et al. 2024). Additionally, China's efforts to digitize public services make it an ideal case for exploring how technology and collaboration can enhance agility. Findings in this study are also applicable to other nations undergoing digital transformations in governance, which may also face parallel issues in achieving agile response by ensuring technological efficiency and smooth collaboration. Based on the results of data analysis and interpretative structural modeling (ISM), we present a holistic theoretical model to explain the process of achieving agile response in e-governance from the perspective of employees. This study contributes to existing knowledge by simultaneously considering the dual pathways, rather than a single path, to achieve agile response. Additionally, it categorizes the antecedents influencing agile response through the dual pathways into five distinct categories using ISM, thereby deepening the understanding of agile response dynamics. Our dual pathways can help characterize and distinguish between the different types of antecedents of agile response and serve as a guide for future research.

**Fig. 1** A dual path to achieve agile response in e-governance

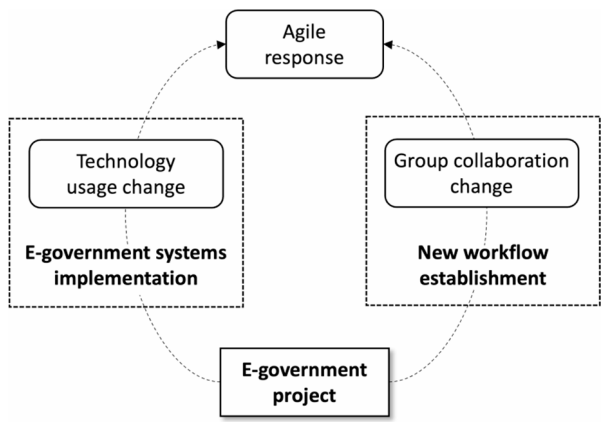

The structure of this study is as follows. The next section reviews existing research on e-government, agile response, technology usage, and group collaboration. Then, we introduce the qualitative methodology and data analysis process. Following that, we summarize the findings and discuss the theoretical contributions, accompanied by practical implications. The final part discusses the limitations and future research directions.

## 2 Literature Review

### 2.1 ICT‑ Driven Government‑to‑Employee

E-government has evolved over approximately two decades, transitioning from internet-based to ICT-driven implementation (Choi and Chandler 2020), and even incorporating artificial intelligence (AI) (Al-Mushayt 2019). Initially, the definitions of e-government stress its technological aspect, suggesting that e-government refers to governments delivering their services via digital means (West 2004).

With the increasing complexity of stakeholder involvement, scholars define e-government by considering the differences in interactions between government organizations and stakeholders. Two crucial patterns are government-to-citizen (G2C) and government-to-employee (G2E) (Gupta et al. 2008; Rao 2011). G2C refers to the interactive relationship between government and citizens, while G2E focuses on internal connections among government employees within government sectors (Rao 2011). For instance, G2C systems enable citizens to access government information and services instantly and conveniently, regardless of their location. This accessibility is facilitated by various channels such as the Internet, email, call centers, kiosks, mobile phones, and other devices. G2E systems are tailored to meet the requirements of government employees, such as online services for tasks such as payroll management, accessing tax information, and accessing resources for human resource training and development, facilitating knowledge sharing and collaboration across various functions. These systems streamline internal workflows and expedite administrative processes, offering efficient solutions for optimizing government operations.

Recently, research has emphasized the holistic transformation driven by e-government, including both technology implementations and workflows of government employees, to enhance service delivery (Choi and Chandler 2020). Collaborative e-government has been emerging in such settings, referring to a new workflow that involves smooth collaborations at different levels in governments (Nam and Pardo 2014). However, limited research has delved into both technology usage and group collaboration from the perspective of government employees.

## 2.2 Agile Response in E-Governance

The overarching goal of e-government is to achieve agile governance by enhancing government effectiveness and efficiency in dealing with specific tasks in volatile contexts (Choi and Chandler 2020). Agile governance refers to the organizational culture and collaborative atmosphere that are flexible, adaptable, and rapid to effectively handle internal tasks and external threats or challenges in government (Mergel et al. 2018). In other words, agile response in governance goes beyond technology usage, indicating a deeper transformation in the workflows (i.e., collaborative atmosphere) among government employees.

Previous studies have reported that e-government project failures are common and can occur at any point during the project lifecycle (Anthopoulos et al. 2016). Meanwhile, a significant amount of existing research has delved into the factors relevant to the success or failure of e-government projects (Anthopoulos et al. 2016; Stefanovic et al. 2016). Therefore, achieving agile response in e-governance has been a subject in prior studies (Mergel et al. 2018). This trend is becoming increasingly prevalent, particularly after the outbreak of COVID-19, because governments are faced with more uncertain and volatile environments to navigate (Janssen and Van Der Voort 2020; Li et al. 2022).

Previous research has primarily focused on the usage of G2C systems, aiming to validate the benefits the systems bring to governments in the public sphere and understand the mechanisms of achieving these benefits (Tolbert and Mossberger 2006; Aladwani and Dwivedi 2018; Arshad and Khurram 2020). Some studies investigate the adoption of G2C systems (Mansoori et al. 2018; Chan et al. 2021; Mirkovski et al. 2023) and attempt to understand the factors influencing their net benefits (Scott et al. 2016). While the citizens' usage is crucial (Lallmahomed et al. 2017), achieving agile response mainly relies on the behaviors of government employees (Stefanovic et al. 2016). However, research on the usage of G2E systems and the specific triggers of collaborative atmosphere driven by G2E systems remains limited.

Other streams of research have explored the paradoxical requirements of agile response, such as the need for high decision speed and accurate analysis results (Janssen and Van Der Voort 2020). From the perspective of communicating with citizens, existing research has revealed the mechanisms through which agile response can be achieved using social media platforms (Li et al. 2022). Research also suggests that clarity regarding the gains and costs of agile work practices is a precondition for the adoption of agile response practices (Mergel 2024). However, despite the crucial role played by government employees in understanding and achieving agile response, fewer studies have considered the mechanisms through which agile response can be achieved from this perspective.

## 2.3 Technology Usage in E-Government

Technology usage is deeply rooted in information systems (IS) research, with numerous theories in this field, including the technology acceptance model (TAM) (Davis 1989), theory of planned behavior (TPB) (Ajzen 1991), IS success model (DeLone and McLean 2003), and the unified theory of acceptance and use of technology (UTAUT) (Venkatesh et al. 2003), which aim to explain users' technology usage intentions and behaviors. Technology usage refers to the utilization of one or more features of a system by a user to perform certain tasks (Andrews 2005). In the context of e-government, government employees use various features of G2E systems to fulfill their daily work tasks. Therefore, the technology usage of government employees is a crucial aspect of understanding the mechanisms of agile response in e-governance.

The mainstream technology usage research has focused on technology adoption and post-adoption stages, considering users' perceptions and behaviors related to the technology (Shaikh and Karjaluoto 2015). For instance, recent studies in the e-government field have explored the determinants of citizens' usage intentions of e-government systems (e.g., Dwivedi et al. 2017) and the spillover effects of their trust in the government (e.g., Porumbescu 2016). Some studies focused on the service design of G2C systems, indicating that citizens' perception of the service design characteristics and service quality influences their overall satisfaction (Chan et al. 2021) and further determines their usage intention (Veeramootoo et al. 2018). Others have emphasized the role of citizens' perceptions and characteristics in influencing their intentions to reuse e-government systems (Mirkovski et al. 2023). Although there have been some initial attempts to understand the failure of e-government (Anthopoulos et al. 2016), including insights from the perspective of government employees regarding the emergence of knowledge vacuums after the implementation of e-government systems (Choi and Chandler 2020), research on how government employees' technology usage contributes to agile response remains limited.

## 2.4 Group Collaboration in E-Government

Collaboration refers to a process involving two or more agents working together towards a shared goal through a set of activities (Viale Pereira et al. 2017). Existing research focused on various aspects of collaboration, including virtual team collaboration (Cheng et al. 2021), and collaborative learning in distributed groups (Cheng et al. 2016b). In the government context, collaboration refers to the sharing of authority or responsibility for decisions on policies, actions, or operations (Harrison et al. 2011). As noted in the evolution of e-government definitions, collaborative e-government has become a prevailing trend (Nam and Pardo 2014). This study primarily focuses on inter-organizational group collaborations among government employees.

Although e-government has garnered attention from scholars, there has been scant research on government employees' group collaborations using G2E systems. With the prevalence of digital transformation, particularly accelerated by the outbreak of COVID-19, governments worldwide have shifted a-lot of their employees work online (Garcia-Rio et al. 2023). By adopting G2E systems, government employees can gain operational benefits, such as reduced paperwork, shorter response times, decreased error rates, improved job outcomes, and providing continuous and agile services to citizens (Akman et al. 2005; Zhang and Venkatesh 2017). Despite these merits, challenges also emerge during the collaboration process. For instance, not all government employees are familiar with, or at ease with collaborations through G2E systems. Adopting technological tools also requires new skills (Stefanovic et al. 2016). Elderly government employees might encounter challenges in using the systems for online collaborations and may perceive online work environments as less trustworthy. As online collaborative working becomes a "new normal", understanding collaborations among government employees and further exploring how it leads to agile response is a priority for the advancement of e-government.

## 3 Research Methodology

### 3.1 Research Background and Methodology Design

The objective of this study is to investigate the dual pathways to achieve agile response in e-governance and find critical factors that lead to agile response in e-governance. Therefore, this exploratory study intends to collect evidence from civil servants directly engaged in the collaborative e-government system. Due to the limited research on the synergy between employees' perception of both technology usage and group collaboration in e-government, exploring its antecedents and mechanisms are of primary importance. An exploratory study based on qualitative research design can help understand complex phenomenon and identify key events and actors in the complicated procedure. Therefore, the exploratory method design is appropriate for an in-depth understanding of complex phenomena.

Specifically, a structured two-stage research approach was used to conduct the data analyzing and theoretical model constructing. Table 1 presents the two-stage research process in this study. In the first stage, following the instructions of Miles and Huberman (1994) on inductive research, we conducted the qualitative data analysis with 2 scholars in our research team. We conducted three-rounds of coding process of the qualitative data and also iterated between existing research and the data to draw conclusions of the exploratory study. In the second stage, we adopted Interpretative Structural Modeling (ISM) to further sort out the path and underlying structure among the factors affecting employees' perception from the first step (Warfield 1974; Jharkharia and Shankar 2005). The interpretive structural model is a method of systems engineering analysis (Guo et al. 2012; Dwivedi et al. 2017). It employs matrix operations and construction to analyze and elucidate the hierarchical relationships and interaction pathways among complex elements within a system. It can effectively explain the essential impact of various elements on targeting variables and investigating the influencing factors in a complex system (Ravi and Shankar 2005; Raj et al. 2008). Due to the lack of systematic analysis of the relationships among various factors contributing to employees' agile response in e-governance, as well as the unclear interaction between technology usage and group collaboration, ISM offers a comprehensive framework and approach to address these challenges.

**Table 1** Two-stage research process

| Stages | Step | Technique adopted | Output |
|---|---|---|---|
| 1. Identifying categories of employees' perception and antecedents of using collaborative e-government. | (1) Interviewing government employees to identify their perceptions of using collaborative e-government | Semi-structured interview | Over 150 pages of qualitative data with 34 participants |
| | (2) Analyzing interview data to identify categories of employees' perceptions in using e-government | Three-rounds of coding | Identified 6 categories and 12 constructs |
| 2. Developing theoretical model of antecedents of trust and reliance in e-government | (3) Developing antecedents of trust and reliance hierarchical framework | Interpretative Structural Modeling (ISM) | Generated a 4-level theoretical model |

**Table 2** Demographics of participants ($N=34$)

| Items | Category | Frequency ($N=34$) | % |
|---|---|---|---|
| Gender | Males | 23 | 67.6 |
| | Females | 10 | 29.4 |
| Age | 25–40 | 15 | 44.1 |
| | 40–60 | 19 | 55.9 |
| Sector | Police | 4 | 11.7 |
| | Construction | 3 | 8.8 |
| | Energy | 7 | 20.6 |
| | Education | 6 | 17.6 |
| | Financial | 3 | 8.8 |
| | Municipality | 6 | 17.6 |
| | Health | 3 | 8.8 |
| | Others | 2 | 5.9 |

## 3.2 Data Collection

We conducted 34 semi-structured interviews to obtain the qualitative data in this study. We chose this sample size based on prior studies using semi-structured interviews (Cheng et al. 2023), which indicate that 30–35 interviews are sufficient for achieving thematic saturation. Also, we noticed that no new themes emerged during the final interviews, indicating the theoretical saturation. Specifically, the target population in this study was civil servants who had experience of online group collaboration using e-government systems. By randomly selecting participants, we sought to minimize potential biases and enhance the generalizability of our findings across different departments in e-government. Table 2 shows the details of the participants who were interviewed. Due to the lockdown, we mainly used personal contacts and con-ducted online interviews through social media, ensuring a 100% response rate. All the participants were informed that the anonymity of their answers and organizations will be guaranteed. The interviews lasted for about 30 min on average depending on the level of details that the participants provided. The interviews were voice-recorded during the interviewing process and then transcribed into a manuscript. The interview material resulted in over 150 pages of qualitative data and was then analyzed as soon as possible after the interview process. The interview protocol consists of three parts: (1) general questions about their background, job responsibility, sector, and department, (2) their perceptions of e-government implication, (3) several open-ended questions about the collaborative e-government working mode.

### 3.3 Stage 1: Coding Process and Interpretation

In this section, we adhered to established methodologies for inductive research to guide our data analysis, as outlined by Miles and Huberman (1994). Two experts from our research team were actively involved in the coding process. The two scholars independently coded the data and then cross-checked their interpretations. To ensure consistency, any discrepancies were discussed and resolved through consensus. Table 3 illustrates this coding procedure in detail. Subsequently, we will provide a thorough explanation of each stage within this coding process.

In the first round of the coding process, we began with general analysis of all the 34 interview transcripts. In particular, two scholars in our research team began with the coding process independently. We went through the transcripts paragraph by paragraph to understand the themes and constructs mentioned by the interviewees and then understand their perceptions of the collaboration e-government during working process. Specifically, we also iterated the interview data with existing research during the coding process. Through the first round of coding, we marked several mentioned themes and extracted corresponding illustrative quotes that conveyed their ideas about the characteristics of using the collaborative e-government system during work time (see the first column in Table 3). For example, participants conveyed their perception of the collaborative e-government process from the perspective of systems, collaboration effectiveness, and other concerns. All the quotes selected were marked with "P(X)", in which X represents the ID of the participants.

In the second round of the qualitative data coding, we extracted more similar and general themes and categories. Based on the illustrative quotes, we classified the extracted relevant variables into different dimensions and categories to help determine the secondary topics (Hoon 2013). In detail, the two researchers conducted the second round of coding and resulted in 12 first-order categories in total, including ease of use, usefulness, traceable, system quality, technology mindfulness, collaboration efficiency, communication transparency, trust in e-government, agile response, and reliance. As shown in Table 3, the first order categories are marked with "AX".

In the third round of the coding, we analyzed the key categories emerging from employees' perspectives on online collaboration in the government sector. Our focus centered on the employees' perception of the e-government system, collaboration quality and the service outcome. Consequently, we identified several second-order core categories in this round: perception of the e-government system, perception of the collaboration quality, perception of the service outcome, trust, reliance, and other concerns. We have denoted the second-order categories with the "AAX".

Then we also provided a summary of the corresponding relationship between each participant and the constructs. The emergent variables of each participant are presented in Table 4.

### 3.4 Stage 2: Interpretative Structural Modeling

Interpretative Structural Modeling (ISM) can provide an efficient way to identify and structure complex problems or backgrounds, which is suitable for our investigation into the antecedents of employees' trust and reliance with collaborative e-government. The main construction steps are as follows: (1) construction of adjacency matrix, (2) construction of reachable matrix, (3) level partition of reachability matrix, (4) establishment of hierarchical structure model.

#### 3.4.1 Construction of Adjacency Matrix

Based on the 10 constructs extracted from the three rounds of coding process in the first stage, further analysis with Interpretative Structural Modeling was then performed by 2 researchers on the identified constructs, to determine the relationship between the two influencing constructs and conduct correlation scores for each pair. The first step of ISM analysis is to establish an Adjacency Matrix. The Adjacency Matrix in this study is presented in Table 5, which act as the input matrix of ISM analysis. Adjacency Matrix is a binary matrix which consists of two elements: the row and column constructs, and the scores of the direct relationship between the constructs in each cell. The constructs are extracted from the coding process in step one and are expected to be refined without redundant and repetitive elements. The score represents the correlation degree between the two constructs. The scoring criteria includes five levels: 0 (no influence), 0.25 (low influence), 0.5 (general influence), 0.75 (high influence) and 1 (significant influence). Specifically, the two interviewers scored the values in the matrix and establish the Adjacency Matrix directectly based on contextual relationships identified from the interview data. According to (Grunert and Grunert 1995), the weak relationships which were mentioned by less than three interviewees in the qualitative data were then removed, so that the hierarchical model will not be too redundant and complex to interpret. After the scoring of each pair of constructs by the two researchers, the final Adjacency Matrix will be determined based on the principle of maximum membership degree. Also, the causality of the element itself does not need to be judged, so the value of the diagonal is fixed at 0. As is shown in Table 5, the final cells in the Adjacency Matrix were populated with value 0 and 1, whereby "$a_{ij}=1$" indicates the direct relationship between construct a and j and "$a_{ij}=0$" indicate otherwise.

#### 3.4.2 Construction of Reachable Matrix

Adjacency matrix is used to describe the direct relationship between the elements, but cannot reflect the indirect relationship between different constructs. Therefore, after

**Table 3** Three rounds of coding process

| Illustrative quotes | First order categories | Second order categories |
|---|---|---|
| Using this system is quite straightforward and uncomplicated, particularly for people in our age group. Overall, navigating through this system usually presents little difficulties. (P4) | A1 Ease of use | AA1 Perception of the e-government system |
| I find navigating the system to be quite user-friendly. Additionally, our company mandates that all employees must be proficient in its use to complete their tasks effectively. (P11) | | |
| During holidays or urgent meetings when gathering in person isn't feasible, software significantly facilitates our communication and collaboration. (P9) | A2 Usefulness | |
| The system's benefit lies in its convenience for instant communication, enabling the transmission of voice and video streams. Additionally, it supports online document management and real-time information processing. (P7) | | |
| In online collaboration, every action is recorded, sometimes even as a document. It's easy to overlook details, but online meetings offer electronic memos and calendars that serve as reminders, ensuring you don't forget important items. (P1) | A3 Traceable | |
| Online platforms create a digital footprint. For instance, if you arrive late to a meeting, you can still access others' notes. (P5) | | |
| I trust the government website. Because the website provided by the government is relatively standardized and has been approved by the national management system, its process and security guarantee are relatively high. (P28) | A4 System quality | |
| The government affairs or official office software you are referring to seems to be thoughtfully designed with stability and functionality in mind. Its overall design appears to effectively align with the demands and positioning of an office environment, providing a user experience that is both efficient and suitable for official tasks. This kind of software design is crucial in ensuring that government work is conducted smoothly and effectively. (P1) | | |
| I'm open to accepting, trying and learning new things (collaborative e-government technology). Many of our colleagues in their 40s and 50s are also willing to learn, and we young people are more willing to accept it (new technology). (P19) | A5 Technology mindfulness | |
| I sometimes want to use the new office software for collaboration. (P22) | | |
| I personally enjoy experimenting with new technologies. I love using new software to improve work efficiency. (P34) | | |

**Table 3** (continued)

| Illustrative quotes | First order categories | Second order categories |
|---|---|---|
| Utilizing video conferencing and related systems can enhance our productivity while also significantly lowering staff turnover, minimizing unnecessary time and resource wastage. (P7) | A6 Collaboration efficiency | AA2 Perception of the collaboration quality |
| Using the online system gives me a greater sense of freedom and can save us travel time. Additionally, it may facilitate smoother communication. (P1) | | |
| Online communication can sometimes lack fluidity, and the interactions between members may not be deep enough. (P10) | | |
| The benefit of online communication lies in its transparency, allowing you to clearly understand your work tasks. Moreover, it tends to be more efficient. (P1) | A7 Communication transparency | |
| Online communication facilitates the immediate editing and sharing of files online, allowing us to understand our colleagues' viewpoints instantly. (P5) | | |
| For instance, upon receiving feedback from the public, we can transmit photos to technicians for a quick assessment of the insects' appearance and the pest-infested areas, enabling the arrangement of prevention and control personnel based on specific conditions. Additionally, the public's needs can be communicated in a group setting, allowing relevant leaders to promptly understand and address these needs. This approach facilitates targeted services and timely responses, ensuring efficient and effective community support. (P2) | A8 Agile response | |
| Leveraging online technology, we ensure immediate responses to any inquiries. We also promptly disseminate health-related information, making our service both convenient and swift. This approach enhances accessibility and efficiency in addressing the public's needs. (P3) | | |
| Since the epidemic, our reliance on online platforms has increased, and my trust in them has grown. Prior to the pandemic, individuals in my age group were less accustomed to utilizing this software. (P4) | A9 Trust in e-government | |
| I trust the system more during the usage. (P6) | | |
| I believe my trust in the software remains strong, despite encountering some issues during use, as it continues to evolve and improve. (P8) | | |
| When I initially began using the system, I was quite perplexed and uncertain. However, over time, I've come to realize that it has become an indispensable part of my workflow. The more I use it, the more my confidence in its capabilities grows. It has now become an integral tool, and it is essential for the progression of departments at every level. This software has effectively automated our work processes, marking a significant achievement. (P10) | A10 Reliance | |
| Our dependence on this software stems from its remarkable convenience. Given the myriad of conveniences it offers, it naturally compels us to use it more frequently. (P29) | | |

**Table 4** Emergent themes by the respondents

| Emergent variables | A1 | A2 | A3 | A4 | A5 | A6 | A7 | A8 | A9 | A10 |
|---|---|---|---|---|---|---|---|---|---|---|
| Participant 1 | | | √ | √ | √ | | | | | |
| Participant 2 | √ | √ | | √ | | √ | | √ | √ | √ |
| Participant 3 | √ | | | | | √ | | | | |
| Participant 4 | √ | | √ | √ | | √ | | √ | | |
| Participant 5 | | | √ | √ | | | | | √ | |
| Participant 6 | | | | √ | | | | | | |
| Participant 7 | | √ | | √ | | | | √ | | |
| Participant 8 | √ | | | | | | | √ | | √ |
| Participant 9 | √ | √ | | √ | | | | | | |
| Participant 10 | √ | √ | | √ | | √ | | | | |
| Participant 11 | √ | √ | | √ | | | | | | |
| Participant 12 | | | | √ | | | | | | √ |
| Participant 13 | | √ | | | | | | | | |
| Participant 14 | | √ | | √ | | | | | | |
| Participant 15 | √ | √ | √ | √ | | √ | √ | | | |
| Participant 16 | √ | √ | | √ | √ | √ | √ | | √ | |
| Participant 17 | √ | √ | | √ | | | √ | | | √ |
| Participant 18 | | √ | | √ | | √ | √ | | | √ |
| Participant 19 | √ | √ | √ | √ | | √ | √ | | | √ |
| Participant 20 | √ | √ | | √ | | √ | √ | | √ | |
| Participant 21 | √ | √ | | √ | √ | √ | √ | | | |
| Participant 22 | √ | √ | | √ | √ | √ | √ | | √ | |
| Participant 23 | √ | √ | √ | √ | | | √ | √ | √ | |
| Participant 24 | √ | √ | | √ | | | | √ | √ | √ |
| Participant 25 | √ | √ | | | | | | | √ | |
| Participant 26 | √ | √ | | √ | | | √ | | | |
| Participant 27 | √ | | | √ | | | | | √ | |
| Participant 28 | √ | √ | | √ | | | | √ | | √ |
| Participant 29 | √ | √ | | √ | | | | √ | | |
| Participant 30 | √ | √ | √ | √ | √ | | √ | √ | | |
| Participant 31 | √ | √ | | √ | √ | | √ | √ | | √ |
| Participant 32 | √ | | | √ | √ | | | √ | √ | √ |
| Participant 33 | √ | | | √ | √ | | | √ | √ | √ |
| Participant 34 | √ | | | √ | | | √ | | | |
| In total | 26 | 23 | 7 | 30 | 8 | 11 | 13 | 12 | 11 | 11 |

obtaining the adjacency matrix of the constructs from the coding process, we used SPSSPRO to transform the input adjacency matrix A into the Reachability Matrix M through matrix operation. Specifically, $I$ represents the identity matrix, $A_i$ represents the Boolean exponentiation matrix, and $i$ is a positive integer. When we obtain the result that $A_1 = A + I$, $A_2 = (A + I)^2$, … $A_i = (A + I)^i$, and $A_1 \neq A_2 \neq A_3 \neq … \neq A_i = A_{i+1} = M$, the reachable matrix $M$ is generated. Table 6 presents the reachable matrix.

**Table 5** Adjacency matrix

| 10 * 10 | A | B | C | D | E | F | G | H | I | J |
|---|---|---|---|---|---|---|---|---|---|---|
| A | 0 | 0 | 0 | 1 | 0 | 1 | 0 | 1 | 0 | 1 |
| B | 0 | 0 | 0 | 0 | 0 | 1 | 0 | 0 | 0 | 0 |
| C | 1 | 1 | 0 | 1 | 0 | 0 | 0 | 1 | 1 | 1 |
| D | 0 | 1 | 0 | 0 | 0 | 0 | 0 | 1 | 0 | 0 |
| E | 1 | 0 | 0 | 0 | 0 | 0 | 0 | 1 | 1 | 1 |
| F | 0 | 0 | 0 | 0 | 0 | 0 | 0 | 0 | 0 | 0 |
| G | 1 | 1 | 0 | 1 | 0 | 1 | 0 | 0 | 1 | 1 |
| H | 0 | 0 | 0 | 0 | 0 | 1 | 0 | 0 | 0 | 0 |
| I | 0 | 1 | 0 | 1 | 0 | 1 | 0 | 1 | 0 | 1 |
| J | 0 | 1 | 0 | 0 | 0 | 0 | 0 | 1 | 0 | 0 |

Note A: System quality, B: Collaboration efficiency, C: Ease of use, D: Technology mindfulness, E: Usefulness, F: Agile response in e-governance, G: Traceable, H: Software reliance, I: Communication transparency, J: Trust

**Table 6** Reachable matrix

| 10 * 10 | A | B | C | D | E | F | G | H | I | J |
|---|---|---|---|---|---|---|---|---|---|---|
| A | 1 | 1 | 0 | 1 | 0 | 1 | 0 | 1 | 0 | 1 |
| B | 0 | 1 | 0 | 0 | 0 | 1 | 0 | 0 | 0 | 0 |
| C | 1 | 1 | 1 | 1 | 0 | 1 | 0 | 1 | 1 | 1 |
| D | 0 | 1 | 0 | 1 | 0 | 1 | 0 | 1 | 0 | 0 |
| E | 1 | 1 | 0 | 1 | 1 | 1 | 0 | 1 | 1 | 1 |
| F | 0 | 0 | 0 | 0 | 0 | 1 | 0 | 0 | 0 | 0 |
| G | 1 | 1 | 0 | 1 | 0 | 1 | 1 | 1 | 1 | 1 |
| H | 0 | 0 | 0 | 0 | 0 | 1 | 0 | 1 | 0 | 0 |
| I | 0 | 1 | 0 | 1 | 0 | 1 | 0 | 1 | 1 | 1 |
| J | 0 | 1 | 0 | 0 | 0 | 1 | 0 | 1 | 0 | 1 |

Note A: System quality, B: Collaboration efficiency, C: Ease of use, D: Technology mindfulness, E: Usefulness, F: Agile response in e-governance, G: Traceable, H: Software reliance, I: Communication transparency, J: Trust

### 3.4.3 Level Partition of Reachability Matrix

Then we conducted the level partition on the Reachability Matrix to determine the ISM hierarchical relationship between various constructs. Before the level partition of the hierarchical relation of elements, the reachability set and the antecedent set as well as their intersection should be defined and analyzed. The reachability set R(Si) is defined as the set of column constructs containing digit 1 in the row construct Si of the Reachability Matrix. The antecedent set A(Si) is defined as the set of row constructs containing digit 1 in the column construct Si of the Reachability Matrix. Intersection set results from $R \cap A$. Based on the above definition, the reachability set R(Si), the antecedent set A(Si) and the intersection set are shown in Table 7.

As is shown in Table 6, the first row of the Reachability Matrix presents that digit 1 occurs for A, B, D, F, H, and J, indicating that S1 can reach the above six constructs. Therefore, the reachability set $R(S1)$ = {A, B, D, F, H, J}. Similarly, as is shown in the first column, digit 1 occurs for the A, C, E, and G, indicating that construct S1

**Table 7** Level Partition of Reachability Matrix

| Iteration | Construct | Reachability set $R(Si)$ | Antecedent set $A(Si)$ | Intersection Set $R \cap A$ | Level |
|---|---|---|---|---|---|
| **1** | A | A, B, D, F, H, J | A, C, E, G | A | L1={F} |
| | B | B | A, B, C, D, E, G, I, J | B | |
| | C | A, B, C, D, F, H, I, J | C | C | |
| | D | B, D, F, H | A, C, D, E, G, I | D | |
| | E | A, B, D, E, F, H, I, J | E | E | |
| | F | F | A, B, C, D, E, F, G, H, I | F | |
| | G | A, B, D, F, G, H, I, J | G | G | |
| | H | F, H | A, C, D, E, G, H, I, J | H | |
| | I | B, D, F, H, I, J | C, E, G, I | I | |
| | J | B, F, H, J | A, C, E, G, I, J | J | |
| **2** | A | A, B, D, H, J | A, C, E, G | A | L2={B, H} |
| | B | B | A, B, C, D, E, G, I, J | B | |
| | C | A, B, C, D, H, I, J | C | C | |
| | D | B, D, H | A, C, D, E, G, I | D | |
| | E | A, B, D, E, H, I, J | E | E | |
| | G | A, B, D, G, H, I, J | G | G | |
| | H | H | A, C, D, E, G, H, I, J | H | |
| | I | B, D, F, H, I, J | C, E, G, I | I | |
| | J | B, H, J | A, C, E, G, I, J | J | |
| **3** | A | A, D, J | A, C, E, G | A | L3={D, J} |
| | C | A, C, D, I, J | C | C | |
| | D | D | A, C, D, E, G, I | D | |
| | E | A, D, E, I, J | E | E | |
| | G | A, D, G, I, J | G | G | |
| | I | D, F, I, J | C, E, G, I | I | |
| | J | J | A, C, E, G, I, J | J | |
| **4** | A | A | A, C, E, G | A | L4={A, I} |
| | C | A, C, I | C | C | |
| | E | A, E, I | E | E | |
| | G | A, G, I | G | G | |
| | I | I | C, E, G, I | I | |
| **5** | C | C | C | C | L5={C, E, G} |
| | E | E | E | E | |
| | G | G | G | G | |

Note A: System quality, B: Collaboration efficiency, C: Ease of use, D: Technology mindfulness, E: Usefulness, F: Agile response in e-governance, G: Traceable, H: Software reliance, I: Communication transparency, J: Trust can be reached by A, C, E, and I. Thus, the antecedent set of construct S1 *A(Si)* = {A, C, E, G}. The intersection set of the reachability set and the antecedent set of S1 results in $R \cap A$ = {A}. The above process was then repeated for all the 10 constructs to complete the first round of iteration of the level partition. The final reachability set, antecedent set, and intersection set for all the construct are presented in Table 7.

Then Table 7 also shows the hierarchical level partition process of the structural model. In each extraction, if the intersection set and reachability set is found to be consistent, the element set can be extracted for hierarchical partition. For example, as shown in the first iteration in Table 7, the construct F has the same reachability set and intersection set, thus resulting in the first hierarchy. Therefore, the construct F is positioned at the first level of the ISM hierarchy. In the following rounds of iteration, the constructs that were extracted as the first level in the previous iterations were deleted, and new constructs were then selected for the successive levels with the same method and process, until all the levels of constructs were finally found. We presented the results of iterations 1–5 in Table 7, respectively.

| Level | Constructs |
|---|---|
| 1 | Agile response in e-governance |
| 2 | Software reliance, Collaboration efficiency |
| 3 | Technology mindfulness, Trust |
| 4 | System quality, Communication transparency |
| 5 | Ease of use, Usefulness, Traceable |

**Table 8** Outcome of the level partition of Reachability Matrix

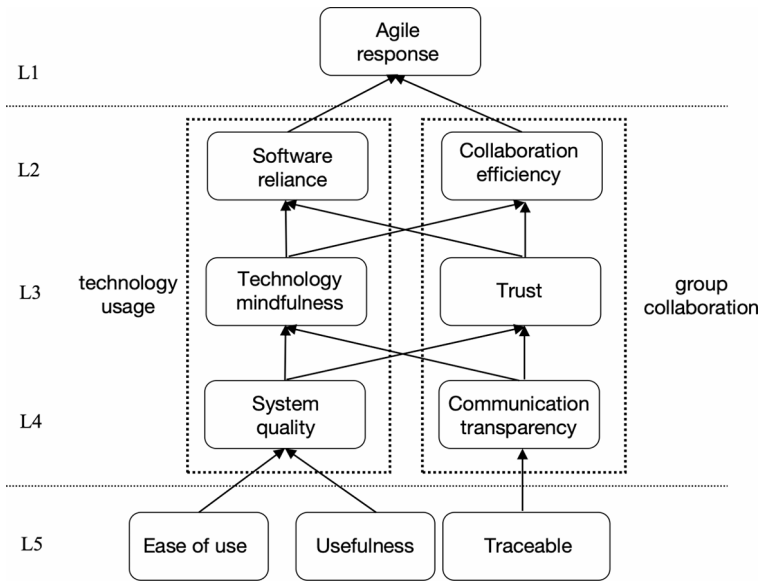

**Fig. 2** Interpretative structural modeling (ISM) model

So far, we have created the adjacency matrices and reachability matrices, followed by the level partition process. As presented in Table 8, the level partition process yields a five-level ISM model. The 10 constructs of this study are divided into 5 hierarchical levels. The first level includes F. The second level includes B and H; The third level includes D and J; The fourth level includes A and I; The fifth level includes C, E, and G. Based on the above results, the interpretive structural model of antecedents of the five-layer mechanisms leading to agile response in e-governance, considering both government employee technology usage and collaboration perspectives is built, as shown in Fig. 2. The diagram depicts the direct relationship between constructs, with arrows representing the direction of each impact. The interpretive structural model illustrates a clear hierarchy of employees' perception and corresponding flow of relationships.

### 3.4.4 Establishment of Hierarchical Structure Model

Drawing from the constructs derived from Stage 1, Fig. 2 presents the ISM model of the dual pathways of both technology usage and group collaboration perception, in which the lower layers are the antecedents of the upper layer. Generally, the ISM model shows that perceived ease of use, usefulness, and traceability of the e-government system are key antecedents of system quality and collaboration efficiency, while perceived high level of system quality and collaboration efficiency affects employees' technology mindfulness and trust, consequently increasing their reliance with the collaborative e-government and collaboration efficiency. Above factors will then contribute to agile response in e-governance. The detailed explanation and theoretical analysis of ISM model results and the dual pathways are as follows.

L5 is positioned in the bottom of the ISM hierarchy, which is the fundamental factor that contributes to the agile response in e-governance, including ease of use, usefulness, and traceable. Ease of use and usefulness are important antecedents and principles for improving technology usage and have been invariably viewed by prior literature (Davis 1989). Traceable are more related to government employees' work flow.

L2, L3, and L4 are the intermediate layer factors, which play a bridging role in the ISM, and are mainly composed of system quality, technology mindfulness, software reliance, communication transparency, trust, and collaboration efficiency. Specifically, system quality, technology mindfulness, and software reliance belong to the government employee technology usage pathway to agile response in e-governance, while communication transparency, trust, and collaboration efficiency belong to the government employee collaboration pathway to agile response.

Constructs of L1 represents the first level and the top of the ISM model, which is agile response in e-governance, indicating the ultimate goal in our theoretical model. Results of the ISM model indicates that software reliance and collaboration efficiency are the direct antecedents to achieve agile response.

*Ease of use, Usefulness, and System Quality* According to the technology acceptance model, perceived ease of use and usefulness of the platform are the primary factors that are driving its usage, thus indirectly influencing the perceived quality of the system (Carter and Bélanger 2005; Lin et al. 2011). Tools and practices that are easy to use have been identified to enhance stakeholder communication and collaboration which are the key aspects of agile response in e-governance. Moreover, while an e-government system deemed useful is more likely to be integrated into employees' daily operations, enhancing its perceived quality due to its ability to meet user needs efficiently and effectively (DeLone and McLean 1992, 2003). For example, when individuals perceive the use of the platform as safe, reliable, convenient and rapid, they will get a better use experience, and their willingness to adopt it will also increase, thus contributing to the agile response in e-governance. However, when users perceive that the e-government system is useless, they will not trust this work-ing paradigm and tend to reduce their willingness to adopt it. Therefore, we conclude that the degree of employees' perception of the usefulness and ease of use of the collaborative e-government system will directly impact their general perceptions, influencing both the utility of the system and agile response (Vlaanderen et al. 2011). Thus, we propose the following proposition.

**Proposition 1** Perceived ease of use and usefulness of the e-government system are positively related to the employees' perceived system quality in the e-government context, laying the foundation for agile response through the pathway of effective technology usage.

*Traceable and Communication Transparency* Traceable indicates that the system provides the function to trace back the actions or records of other employees (Davison 2020; Sipior 2020). In contrast to traditional face-to-face collaboration between employees, online collaboration demands a more transparent exchange of information that is accurate, traceable, and expedient (Zhao et al. 2012). System Traceability facilitates a clear, accessible record of decisions, changes, and working progress. This concept also emphasizes the extent to which group members are open with each other. Digital platforms facilitate this process by ensuring that communication remains transparent. It reduces uncertainties and ambiguities, allowing employees to have a shared understanding of objectives, challenges, and the rationales behind decisions. Therefore, the traceability of the e-government system will also impact employees' perceptions and is fundamental to achieving communication transparency. Thus, we propose the following proposition.

**Proposition 2** Perceived traceable of the e-government system are positively related to the communication transparency in the e-government context, laying the foundation for agile response through the pathway of group collaboration of employees.

*System Quality, Technology Mindfulness, and Software Reliance* The important attributes that form a high level of perceived system quality of the employees in this study are usefulness and ease of use in the L5. Technology mindfulness refers to users' willingness to explore new features with a system or consider more applications in the specific kind of system (Thatcher et al. 2018). Individuals with high level of technology mindfulness usually demonstrate curiosity, flexibility, and experimentation when interacting with the systems. Taking a communication system as an example, less mindful users tend to rely on the familiar features to communicate, such as the "voice" function, while more mindful users like to experiment with more functions or features to increase productivity, such as the screen sharing and virtual reality. A high level of technology mindfulness indicates an environment that encourages employees' reflective engagement with technology, thus enhancing their reliance on the system (Swanson and Ramiller 2004). High quality systems will enhance user willingness to engage with them, consequently fostering a positive intention to experience more functions of the e-government systems. Enhancing the efficiency of subjective encounters can also significantly diminish the likelihood of behavioral transference, thus improving their reliance to the collaborative e-government system. Thus, we propose the following proposition.

**Proposition 3** The perceived high quality of the e-government system leads to a thoughtful engagement and fosters a preference and appreciation with the system capability, which is for cultivating technology mindfulness and subsequent reliance on the e-government systems. Technology mindfulness and software reliance act as a mediating role in the "technology usage" pathway of achieving agile response in e-governance.

**Communication Transparency, Trust, and Collaboration Efficiency** Trust in this research context is mainly characterized by employees' subjective assessment of the group collaboration and perceived outcomes when experiencing the collaboration process. It constitutes a comprehensive entity that encompasses interpersonal trust between the employees during the working flow. Findings of this study support that employees' trust depends on the communication transparency between employees. Communication transparency enhance trust by reducing misunderstandings, fostering positive interactions, and ensuring shared understanding, which fosters mutual trust (Schoop et al. 2014). A fundamental aspect of trust is the inclination of individuals to rely on the words, actions, or decisions of others (Lowry et al. 2010). Effective collaboration, characterized by substantial interaction, prompts individuals to rely on each other's opinions. Consequently, a high level of trust among individuals serves as a catalyst for building a high-quality collaboration process (Cheng et al. 2016b). Thus, the following proposition was proposed.

**Proposition 4** A high level of communication transparency during the group collaboration fosters an environment of employees' trust, which in turn enhances employees' collaboration efficiency. Trust and collaboration efficiency act as a mediating role in the "group collaboration" pathway of achieving agile response in e-governance.

*Software Reliance, Collaboration Efficiency, and Agile Response in E-Governance* In this research context, agile response is the ultimate goal of collaborative e-governance, which refers to the government sectors' ability to efficiently adapt to and incorporate changes in requirements during service delivery (Lee and Xia 2010). According to the ISM model, the dual pathways to achieve agile response in e-governance consists of not only technology usage but also employees' perceptions during the group collaboration process. On one hand, the technological pathway emphasizes the achievement of governmental and organizational objectives through technological empowerment, involving employees effectively engaging with the features of the e-government system to complete tasks (Andrews 2005). On the other hand, the perception of collaborative effects focuses more on the openness and trustworthiness of the collaborative environment at the perceptual level. Thus, the following proposition is proposed.

**Proposition 5** A high level of software reliance and employee collaboration efficiency enhances agile response in e-governance, facilitated by the pathways of technology usage and group collaboration, respectively.

## 4 Discussion and Conclusion

### 4.1 Findings

For government employees, the path to achieve agile response is characterized by the synergy of dual pathways, rather than a singular trajectory, encompassing both technology usage and collaboration between the employees. This study explored the dual pathways to achieve agile response in e-governance through qualitative analysis. Using Interpretative Structural Modeling, we presented the underlying structure between different factors in the e-government background and developed a hierarchical model. The findings provide strategic guidance for both the government sector to encourage collaboration efficiency, and enhance the design of the e-government systems, thus improving employees' efficiency and effectiveness of e-government service providing and enabling the agile response of the government sector.

Based on the findings of our research, we asserted that the dual pathways of agile response in e-governance in the Government-to-Employee (G2E) context mainly derived from the synergy between technology usage and collaboration. Antecedents of the two pathways consist of multifaceted constructs encompassing five hierarchical levels: the fifth level consists of ease of use, usefulness, and traceability; the fourth level consists of system quality and communication transparency; the third level consists of technology mindfulness and trust; the second level consists of software reliance and collaborative efficiency; and finally the first level consists of agile response in e-governance. Regarding the technology usage pathway, perceived usefulness and ease of use synergistically contribute to the perceived system quality, which subsequently augments technology mindfulness among employees. This augmentation in technology mindfulness will then cultivate a higher reliance on the e-government system, thereby facilitating agile response. Regarding the collaboration pathway, the augmentation of system traceability significantly enhances perceived communication transparency during the collaboration process, thereby facilitating trust among employees. A high level of trust will then catalyze the efficiency of the collaborative processes, creating a conducive environment for agile response.

### 4.2 Theoretical Contributions

Delivering satisfactory public services is a fundamental responsibility of the government. Leveraging online channels enhances efficiency and transparency in service delivery and communication with the public. To gain the satisfaction and trust of the citizens, the government must prioritize ensuring that its own employees perceive e-government systems as of the highest quality. Employee endorsement is pivotal in achieving agile response in e-governance (Janita and Miranda 2018).

Regarding this, the first contribution of this study lies in its proposal of the dual pathways and essential antecedents that an e-government system should encompass for the employees to achieve agile response. Despite the existing knowledge on technology usage and collaboration, limitations persist in adequately explaining the synergy between these elements in realizing agile response from the perspective of government employees, especially within the e-government context (Nam and Pardo 2014; Choi and Chandler 2020). Our research concentrates on the G2E context, specifically examining websites where the employees interact with the e-government systems or other employees. This will guide researchers in this field to develop more theoretical framework and test some new hypotheses to explain the synergy between different mechanisms in achieving agile response in e-governance.

The second theoretical contribution pertains to classifying the antecedents of achieving agile response into five categories, with the help of Interpretative Structural Modeling. These categories can provide suggestions from several perspectives, including both system characteristics, and collaborations between employees. Findings emphasize the necessity for designers and developers to prioritize these aspects to enhance employees' positive perception and overall system effectiveness. The ongoing exploration and validation of these relationships continue to enrich the field, offering insights into the complex dynamics between group collaboration and technology success. Therefore, the interconnections among the antecedents that bring about the complex theoretical framework have been explored in this study. This theoretical framework provides valuable and holistic insights into the underlying structure of e-government perceptions and the hierarchical model among them. The classification and complex structure help us enhance the understanding of factors that influence successful collaboration between employees and systems or other employees in public sector.

### 4.3 Practical Implications

This study also presents significant practical implications for the successful implementation and collaboration of employees in government sector. First, this study equips decision-makers in government sectors with a comprehensive understanding of the specific factors that will enhance employees' perception of the collaborative e-government. These insights help the leaders to skillfully navigate the complexities of successful adoption of digital technology in the e-government setting. The hierarchical framework in this study can also aid decision-makers in comprehending the multi-layered factors affecting employees' perception. Utilizing this framework, they can also effectively reduce the risk of failures during the e-government implementation phase.

Secondly, the study offers invaluable perspectives for e-government system developers, enabling them to develop the e-government solutions and systems that are directly aligned with the requirements of government decision-makers and employees. It highlights the criticality of ensuring the key features of e-government systems, including the usefulness, ease of use, traceability and the enabling of collaboration efficiency. These elements are vital for the employees' perceived system quality and efficient deployment of e-government. By integrating these insights, develop-ers can produce e-government systems that significantly enhance the probability of successful technology adoption in the government domain, thus increasing employees' positive perceptions with e-government from both technological perspective and collaboration perspective, thus delivering satisfactory public services and achieving agile response.

Moreover, this study also provides suggestions for employees in the government sector regarding e-government system use, especially highlighting the importance of communication transparency between employees. By ensuring communication transparency and efficiency, employees can provide agile responses to the citizens' needs and provide high quality government service. The manager in the government sector can ensure a transparent flow of information, which is essential in fostering a positive atmosphere of collaboration.

## 5 Limitations and Future Research

There are also limitations in this study. First, we attempted to investigate the employees' perceptions of collaborative e-government in general with a hierarchical framework. Different cultural backgrounds or demographic characteristics of employees were not considered in this research. Also, we randomly selected participants to minimize potential biases, while this approach may result in less focus on specific e-government sectors or job-specific characteristics. Therefore, future research can be extended by contextualizing the framework in different countries or considering employees' characteristics, such as job titles and leadership roles, to have a better understanding of the influence of hierarchical roles on collaboration. Purposive sampling or specific analysis can also help to provide deeper insights into how e-governance achieves agile response in the future research. Second, this study proposes the theoretical framework by using a qualitative research design. Future studies can extend the work by developing new hypotheses and collecting quantitative data to test them empirically, which can be complementary to the conclusion of this study. Finally, future research can target a specific technology adoption such as Generative Artificial Intelligence technologies (e.g. ChatGPT) in the government or economy setting.

**Acknowledgements** This research would like to thank the National Natural Science Foundation of China (Grant No. 72401293, 72434006, 72271236), the Key Projects of Philosophy and Social Sciences Research of Chinese Ministry of Education (Grant No. 23JZD022), the China Postdoctoral Science Foundation (Grant No. 2022M723491), the Science Foundation of China University of Petroleum, Beijing (Grant No. 2462023SZBH006), the Blockchain Lab at the School of Information at Renmin University of China, and Public Computing Cloud in Renmin University of China for providing support for part of this research.

**Data Availability** The data are available from the corresponding author on reasonable request.

**Competing Interests** None